\documentclass[aps,pra,reprint,amsmath,amssymb,superscriptaddress,nofootinbib,longbibliography]{revtex4-2} 
\usepackage{graphicx}
\usepackage[utf8]{inputenc}
\usepackage[T1]{fontenc}
\usepackage{xcolor}
\usepackage{soul}

\usepackage{booktabs}

\usepackage{amssymb}

\usepackage{bm}
\usepackage{natbib}
\usepackage{mathtools}

\begin{document}

	\title{Characterization of magnetic properties, including magnetocaloric effect, of RE$_{5}$Pt$_2$In$_4$ (RE = Gd-Tm) compounds}

	\author{Altifani Rizky Hayyu}
	\email{altifani.hayyu@doctoral.uj.edu.pl}
	\affiliation{M.~Smoluchowski Institute of Physics, Jagiellonian University,
		prof. Stanis\l{}awa \L{}ojasiewicza 11, PL-30-348 Krak\'ow, Poland}
	\author{Stanis\l{}aw Baran}
	\email{stanislaw.baran@uj.edu.pl}
	\affiliation{M.~Smoluchowski Institute of Physics, Jagiellonian University,
		prof. Stanis\l{}awa \L{}ojasiewicza 11, PL-30-348 Krak\'ow, Poland}
	\author{Andrzej Szytu\l{}a}
	\affiliation{M.~Smoluchowski Institute of Physics, Jagiellonian University,
		prof. Stanis\l{}awa \L{}ojasiewicza 11, PL-30-348 Krak\'ow, Poland}

\date{\today}
	
\begin{abstract}

The RE$_{5}$Pt$_2$In$_4$ (RE = Gd-Tm) rare earth compounds have been investigated by means of X-ray diffraction (XRD)
as well as by DC and AC magnetometric measurements. The compounds crystallize in an orthorhombic crystal structure
of the Lu$_{5}$Ni$_2$In$_4$-type (\textit{Pbam} space group, No.~55). With decreasing temperature the intermetallics
undergo a transition from para- to ferro-/ferri- (RE~=~Gd, Tb, Ho, Er) or antiferromagnetic state (RE~=~Tm).
In case of Dy$_{5}$Pt$_2$In$_4$, the ferromagnetic state is reached through an intermediate antiferromagnetic order
present in a limited temperature range. The critical temperatures of magnetic ordering range from 4.1~K (RE~=~Tm)
up to 108~K (RE~=~Tb). For the majority of investigated compounds, a cascade of additional magnetic transitions
is found below the respective critical temperatures of magnetic ordering. The magnetic moments are found solely
on the rare earth atoms, while the moments of the remaining Pt and In atoms are absent or are too small to be detected
while accompanied by the strong rare earth's moments. The magnetocaloric (MCE) performance of RE$_{5}$T$_2$In$_4$
(RE~=~Gd-Tm) is found quite good, especially while taking into account the compounds with RE = Ho and Er. Maximum
magnetic entropy change ($-\Delta S_M^{max}$) reaches 11.8 (RE~=~Ho) or 11.4~J$\cdot$kg$^{-1}\cdot$K$^{-1}$ (RE~=~Er)
under magnetic flux density change of 0-9~T. Under the same conditions, the relative cooling power (RCP) and
refrigerant capacity (RC) equal 607 and 495~J$\cdot$kg$^{-1}$ (RE~=~Ho) or 434 and 341~J$\cdot$kg$^{-1}$ (RE~=~Er).

		\bigskip
		
		\noindent \textbf{keywords}: rare earth intermetallics, magnetic properties, magnetization, magnetocaloric effect, magnetic entropy change, relative cooling power
		
	\end{abstract}
	
	\maketitle
	
	\section{Introduction}
	\label{intro}
	
	Rare earth intermetallic compounds have been attracting researchers' interest due to a number of intriguing physical phenomena including multiple magnetic transitions, metamagnetism, large magnetocaloric effect (MCE),
	spin glass state, heavy fermion behavior, superconductivity, and many more. A review paper by Gupta and Suresh~\cite{GUPTA2015562},
	reporting the current state of knowledge for the RTX (R = rare earths, T = Sc, Ti, Mn, Fe, Co, Ni, Cu, Ru, Rh, Pd, Ag, Os, Ir,
	Pt, Au, and X = Al, Ga, In, Si, Ge, Sn, As, Sb, Bi) series of intermetallics, may provide an idea of the multitude of
	interesting phenomena observed in rare earth intermetallic compounds.
	
	Nowadays, the compounds with complex crystal structure and magnetic properties resulting from multiple magnetic sublattices are of particular interest.
	The RE$_{5}$Pt$_2$In$_4$ (RE = Gd-Tm) indides are a good example of such a family of compounds, as they crystallize in an
	orthorhombic structure of the Lu$_{5}$Ni$_2$In$_4$-type (\textit{Pbam} space group, No. 55)~\cite{zaremba2007}.
	In this structure, reported for the first time by Zaremba et al.~\cite{zaremba1991crystal}, the rare earth atoms occupy
	three different Wyckoff sites, namely, the 2(a) site and two 4(g) sites with different atomic positional parameters.
	
	Up to now magnetic properties of RE$_{5}$Pt$_2$In$_4$ (RE = Gd-Tm) remain unexplored. However, such properties including magnetic structures, have been reported for the isostructural
	RE$_5$Ni$_2$In$_4$~\cite{tyvanchuk2010magnetic,provino2012crystal,GONDEK201210,szytula2013magnetic,SZYTULA2014149,Ritter_2015,zhang2018investigation}
	and RE$_5$Pd$_2$In$_4$~\cite{Baran2021} (RE = Tb-Tm). Both the Ni- and Pd-based intermetallics are found to order magnetically at
	low temperatures with the critical temperatures of magnetic ordering ranging from about 4~K in the Tm-based compounds up to 125~K
	reported for Tb$_{5}$Ni$_2$In$_4$. The magnetic moments are carried solely by the rare earth atoms (Ni, Pd and In remain non-magnetic
	or have magnetic too small to be detectable in presence of the strong rare earth moments).
	Below the respective critical temperature, a cascade of temperature-induced magnetic transitions
	is observed for most of the compounds. Neutron diffraction data reveal that these complex magnetic properties can be attributed
	to either different ordering temperatures in different rare earth sublattices or order-order magnetic transitions.
	The low-temperature magnetic structures show a variety of forms including ferro- and antiferromagnetic spin arrangements,
	as well as the coexistence of both types of ordering in selected compounds. The reported magnetic structures include
	both commensurate and incommensurate ones.
	
	Magnetocaloric effect (MCE) in the rare earth-based compounds is recently of great interest due to possible applications in refrigeration~\cite{LI20231,Guo2022,XU2022118114,ZHANG2022100786,MA2021138,ZHANG20191173,LI2020153810,ZHANG2022117669,XU2021100470,LI2020354,ZHANG202266}. It is until now that MCE in the RE$_{5}$T$_2$In$_4$ (RE = rare earth element; T = transition metal) intermetallics has only been reported for RE$_{5}$Ni$_2$In$_4$ (RE = Dy, Ho, and Er)~\cite{zhang2018investigation}. It has been found that for RE = Ho and Er, the maximum entropy changes exceed 10~J$\cdot$kg$^{-1}\cdot$K$^{-1}$ under magnetic flux change 0-7~T. The maximum entropy changes are reached in the vicinity of the respective Curie temperatures, which are close to 20~K.
	
	Complex magnetic properties, reported for the isostructural RE$_5$Ni$_2$In$_4$ and RE$_5$Pd$_2$In$_4$ (RE = Tb-Tm), has inspired
	us to undertake the current study concentrated on RE$_{5}$Pt$_2$In$_4$ (RE = Gd-Tm). We report here not only the basic magnetic
	properties, like magnetic transition temperatures, effective magnetic moments, the moments under applied magnetic field,
	critical and coercivity fields, etc., but also we report the magnetocaloric effect studied under magnetic flux change up to 0-9~T.
	As a result, we determine the number of parameters important from the point of view of potential application in a low-temperature
	refrigeration. The parameters include temperature averaged magnetic entropy change (TEC), relative cooling power (RCP), and
	refrigerant capacity (RC). The experimental results reported in this work are compared with those reported previously for the
	isostructural RE$_5$Ni$_2$In$_4$ and RE$_5$Pd$_2$In$_4$ (RE = Tb-Tm).
	
	\section{Materials and methods}
	
	The samples have been prepared by arc melting of high-purity elements (at least 99.9 wt \%) under titanium-gettered
	argon atmosphere. The elements have been taken in a stoichiometric ratio. The obtained ingots have been remelted
	a few times in order to improve their homogeneity. No annealing have been applied as our previous tests have shown that
	annealing leads to the appearance of impurity phases. The crystal structure of the obtained samples has been examined by
	X-ray powder diffraction at room temperature using a PANalytical X'Pert PRO diffractometer (Cu K$\alpha$ radiation,
	Bragg-Brentano geometry, measured angle interval of $2\theta = 10-100^\circ$, $2\theta$ step = 0.033$^\circ$,
	150~s/step). The X-ray diffraction data have been processed using the FullProf program
	package~\cite{Rodriguez-Carvajal_Physica_B_192,Rodriguez-Carvajal_Newsletter_26}.
	
	For the DC and AC magnetic measurements, the powder samples have been encapsuled in plastic containers and glued by
	varnish in order to prevent grain rotation or movement during the measurement. A Vibrating Sample Magnetometer (VSM)
	option of the Physical Properties Measurement System (PPMS) by Quantum Design has been utilized for DC magnetic measurements.
	Magnetic susceptibility data have been collected over a wide temperature range of 1.85--390~K using both the
	Zero Field Cooling (ZFC) and Field Cooling (FC) regimes. Every time before collecting a ZFC curve, the sample has been
	heated to a temperature exceeding the respective critical temperature of magnetic ordering, then demagnetized by the oscillating
	magnetic field, and finally cooled down to 1.9~K. The ZFC and FC collected at a low magnetic field of 50~Oe have been
	used to determine magnetic transition temperatures, while from the ZFC data taken at 1~kOe the effective magnetic moments ($\mu_{eff}$)
	and paramagnetic Curie temperatures ($\theta_p$) have been derived. In order to follow the thermal evolution of magnetic order,
	isothermal magnetization curves have been collected in magnetic fields up to 90~kOe (9~T) at a number of selected temperatures.
	Before collecting a magnetization curve, the sample has been demagnetized using the described above procedure.
	
	AC magnetic susceptibility measurements have been performed using an AC Measurement System (ACMS) option of PPMS.
	The data have been collected under an oscillating magnetic field of 2~Oe amplitude at a number of selected frequencies
	between 100 and 5000~Hz. The temperature interval
	has covered 1.9--300~K.
	
	The crystal structure has been visualized (see Fig.~\ref{fig:The_crystal_structure_of_RE$_{5}$Pt$_2$In$_4$}
	with the use of the VESTA program~\cite{Momma:db5098}.
	
	\section{Results}
	\subsection{Crystal structure}
	
	\begin{figure}[ht!]
		\centering
		\includegraphics[width=0.50\textwidth]{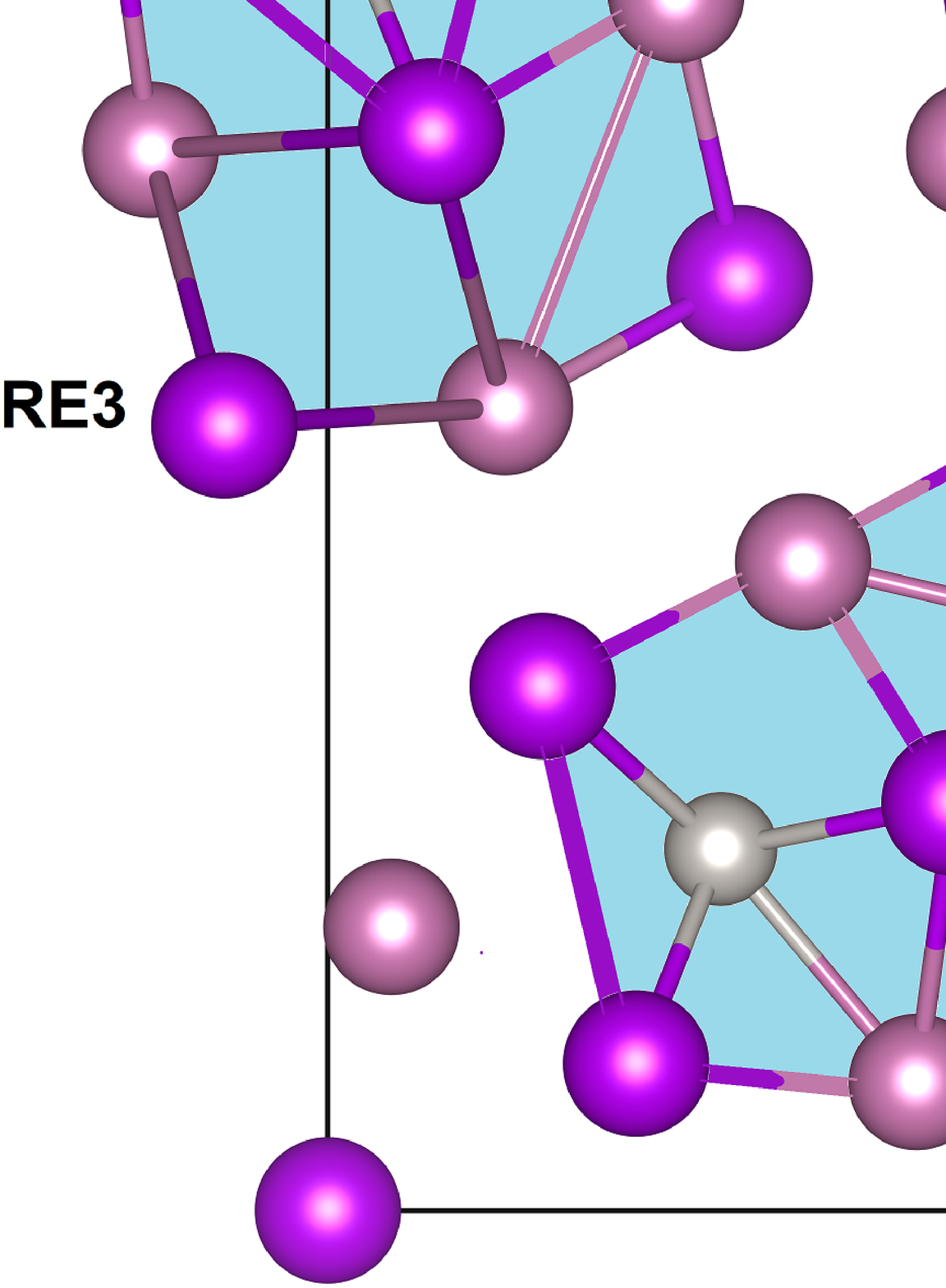}
		\caption{\label{fig:The_crystal_structure_of_RE$_{5}$Pt$_2$In$_4$}Perspective view of the crystal structure of RE$_{5}$Pt$_2$In$_4$ showing its layered nature.}
	\end{figure}
	
	X-ray powder diffraction data collected for the RE$_{5}$Pt$_2$In$_4$ (RE = Gd-Tm) samples at room temperature
	confirm that the compounds crystallize in an orthorhombic crystal structure of the Lu$_{5}$Ni$_2$In$_4$-type
	(Pearson symbol oP22, space group \textit{Pbam}, No. 55, $Z = 2$). This result is in agreement
	with the previous report~\cite{zaremba2007}. The Lu$_{5}$Ni$_2$In$_4$-type structure consists of
	distorted trigonal and square prismatic slabs of the REPt and REIn compositions~\cite{zaremba1991crystal}.
	The structure is a layered-one with atoms situated on the mirror planes at $z = 0$ and $z = \frac{1}{2}$.
	The layers formed by the rare earth atoms are separated by the layers containing Pt and In
	(see Fig.~\ref{fig:The_crystal_structure_of_RE$_{5}$Pt$_2$In$_4$}). The rare earth atoms occupy
	three different Wyckoff sites, namely, one 2a site (0,0,0) and two 4g sites $(x,y,0)$ with different atom
	positional parameters. The Pt atoms occupy one 4h site $(x,y,\frac{1}{2})$, while the In atoms are located at two other 4h sites
	with different atom positional parameters.

	\begin{figure*}
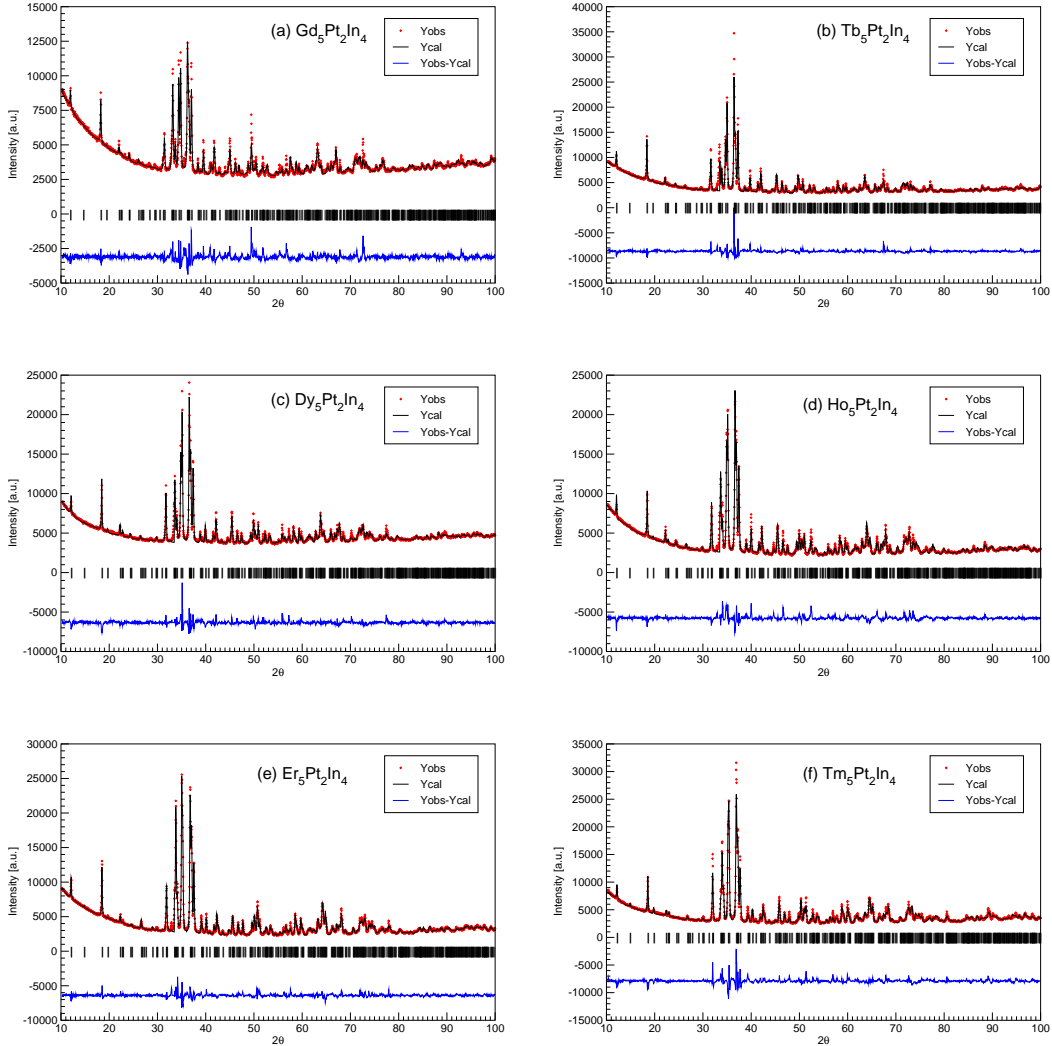

		\centering
		\includegraphics[scale=0.25, bb=8 21 818 573]{Gd5Pt2In4_non-annealed_no_2021_03_XRD.eps}
		\includegraphics[scale=0.25, bb=8 21 818 573]{Tb5Pt2In4_non-annealed_no_2021_02_XRD.eps}
		\includegraphics[scale=0.25, bb=8 21 818 573]{Dy5Pt2In4_non-annealed_no_2021_06_XRD.eps}
		\includegraphics[scale=0.25, bb=8 21 818 573]{Ho5Pt2In4_non-annealed_no_2021_07_XRD.eps}
		\includegraphics[scale=0.25, bb=8 21 818 573]{Er5Pt2In4_non-annealed_no_2021_13_XRD.eps}
		\includegraphics[scale=0.25, bb=8 21 818 573]{Tm5Pt2In4_non-annealed_no_2021_09_XRD.eps}
		\caption{\label{fig:XRD}X-ray diffraction patterns collected at room temperature for RE$_{5}$Pt$_2$In$_4$: (a) RE = Gd, (b) RE = Tb,
			(c) RE = Dy (d) RE = Ho, (e) RE = Er and (f) RE = Tm. The solid circles denote the
			experimental data, while the black lines show the Rietveld refinement results.
			The vertical bars indicate Bragg reflection positions, whereas the difference between
			the experimental data and the refinement is presented as the blue line at the bottom
			of each subfigure.}
	\end{figure*}

	\begin{table*} 
		\begin{footnotesize}
			\begin{flushleft}
				\normalsize 
				\caption{\label{tbl:crystallographic_data}
					\ Crystallographic data obtained from Rietveld refinement of the X-ray powder diffraction patterns collected
					at room temperature for RE$_{5}$Pt$_2$In$_4$ (RE = Gd-Tm; Lu$_{5}$Ni$_2$In$_4$-type structure, space group $Pbam$, No. 
					55). The agreement factors R$_{profile}$, R$_{F}$, R$_{Bragg}$, and $\chi^{2}$ characterizing quality of refinements
					are listed at the bottom of the table.}
			\end{flushleft}
			\vspace{-0.2 cm}
			\footnotesize
			\begin{tabular*}{0.99\textwidth}{@{\extracolsep{\fill}}lllllll}
				\hline
				RE & Gd & Tb & Dy & Ho & Er & Tm\\
				\hline
				$a$ [\r{A}] & 18.2045(20) & 18.1218(14) & 18.0551(13) & 17.9991(15) & 17.9447(11) & 17.8935(18)\\
				$b$ [\r{A}] & 8.0461(9) & 8.0140(6) & 7.9886(5) & 7.9731(6) & 7.9524(5) & 7.9212(8)\\
				$c$ [\r{A}] & 3.6771(5) & 3.6500(5) & 3.6270(3) & 3.6115(4) & 3.5942(2) & 3.5776(5)\\
				$V$ [\r{A}$^3$] & 538.61(23) & 530.08(16) & 523.13(14) & 518.28(16) & 512.92(12) & 507.08(20)\\
				RE1 at 2a (0, 0, 0) & 0 & 0 & 0 & 0 & 0 & 0\\
				RE2 at 4g (x, y, 0) & $x=0.2181(10)$ & 0.2192(8) & 0.2200(7) & 0.2180(9) & 0.2190(7) & 0.2184(8)\\ 
				& $y=0.2394(28)$ & 0.2459(21) & 0.2399(19) & 0.2389(24) & 0.2470(18) & 0.2471(23)\\
				RE3 at 4g (x, y, 0) & $x=0.4204(12)$ & 0.4181(10) & 0.4168(8) & 0.4215(11) & 0.4168(7) & 0.4181(9)\\
				& $y=0.1238(21)$ & 0.1215(17) & 0.1150(15) & 0.1196(19) & 0.1244(14) & 0.1218(18)\\
				Pt at 4h (x, y, $\frac{1}{2}$) & $x=0.3056(8)$ & 0.3031(6) & 0.3028(5) & 0.3043(6) & 0.3044(5) & 0.3012(8)\\ 
				& $y=0.0217(20)$ & 0.0236(13) & 0.0265(12) & 0.0259(14) & 0.0257(12) & 0.0295(17)\\
				In1 at 4h (x, y, $\frac{1}{2}$) & $x=0.5695(14)$ & 0.5678(9) & 0.5645(8) & 0.5652(10) & 0.5691(9) & 0.5606(11)\\ 
				& $y=0.2152(26)$ & 0.2102(17) & 0.2175(17) & 0.2111(17) & 0.2056(15) & 0.2094(19)\\
				In2 at 4h (x, y, $\frac{1}{2}$) & $x=0.8459(14)$ & 0.8494(10) & 0.8492(9) & 0.8511(11) & 0.8484(8) & 0.8516(12)\\ 
				& $y=0.0753(27)$ & 0.0759(18) & 0.0748(18) & 0.0714(20) & 0.0778(18) & 0.0759(24)\\
				R$_{profile}$ [\%] & 2.76 & 3.07 & 2.40 & 3.59 & 3.07 & 4.39\\ 
				R$_{F}$ [\%] & 8.19 & 9.55 & 6.08 & 6.71 & 4.88 & 5.95\\ 
				R$_{Bragg}$ [\%] & 12.5 & 11.3 & 9.34 & 10.9 & 6.96 & 9.70\\ 
				$\chi^{2}$ & 7.30 & 6.87 & 6.46 & 10.4 & 8.14 & 16.3\\
				\hline
			\end{tabular*}
		\end{footnotesize}
	\end{table*}
	
	Fig.~\ref{fig:XRD} shows X-ray powder diffraction patterns of RE$_{5}$Pt$_2$In$_4$ (RE = Gd-Tm) collected
	at room temperature, together with the results of Rietveld refinement. The refined parameters of the crystal
	structure are listed in Table~\ref{tbl:crystallographic_data}.
	
	\begin{figure*}
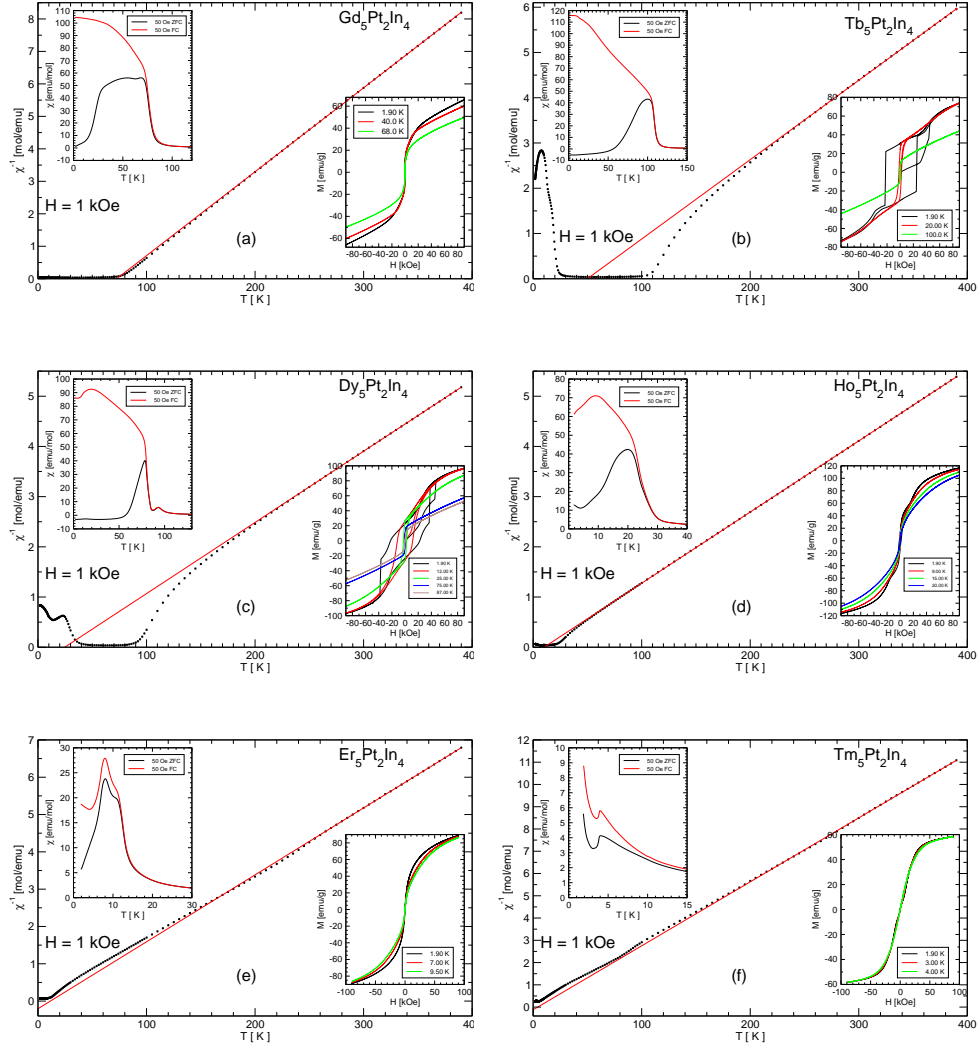

		\centering 
		\includegraphics[scale=0.25, bb=16 21 750 573]{Gd5Pt2In4_non-annealed_no_2021_03_powder_in_varnish_susc.eps}
		\includegraphics[scale=0.25, bb=16 21 750 573]{Tb5Pt2In4_non-annealed_no_2021_02_powder_in_varnish_susc.eps}
		\includegraphics[scale=0.25, bb=16 21 750 573]{Dy5Pt2In4_non-annealed_sample_2021_06_magn_recip_susc.eps}
		\includegraphics[scale=0.25, bb=16 21 750 573]{Ho5Pt2In4_non-annealed_sample_2021_07_magn_recip_susc.eps}
		\includegraphics[scale=0.25, bb=16 21 750 573]{Er5Pt2In4_non-annealed_no_2021_13_powder_in_varnish_susc.eps}
		\includegraphics[scale=0.25, bb=16 21 750 573]{Tm5Pt2In4_non-annealed_no_2022_17_powder_in_varnish_susc.eps}
		\caption{\label{fig:Reciprocal_X_vs_T}Reciprocal magnetic susceptibility with the fitted line represents the Curie-Weiss law for
		RE$_{5}$Pt$_2$In$_4$: (a) RE = Gd, (b) RE = Tb, (c) RE = Dy (d) RE = Ho, (e) RE = Er, and (f) RE = Tm.
		The low-temperature behavior measured at 50~Oe (ZFC and FC regimes) is shown in
		the upper insets, while the isothermal magnetization vs. external magnetic field at selected temperatures
		is presented in the lower insets.}
		\end{figure*}
	
	\begin{figure*}
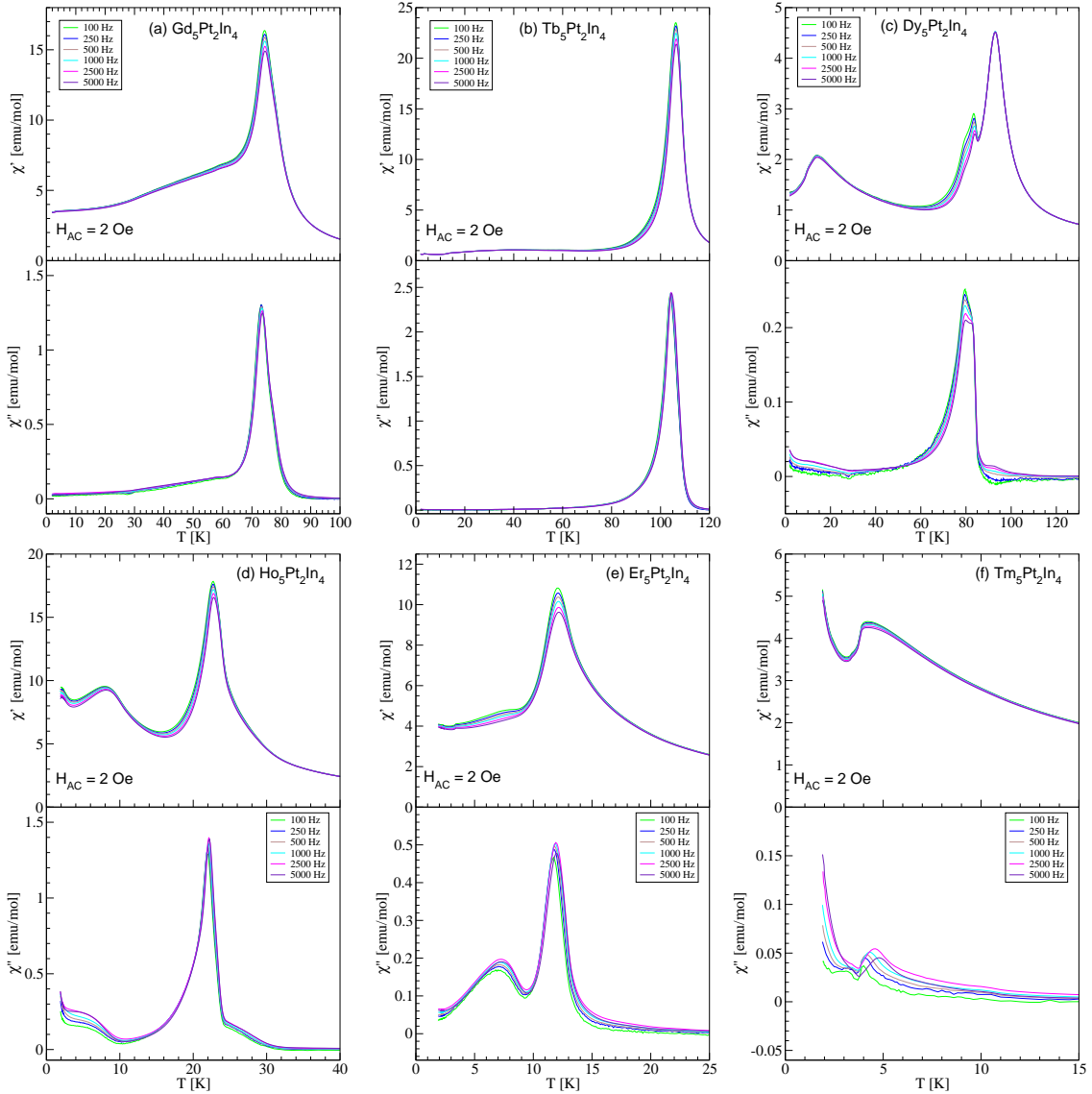

		\centering
		\includegraphics[scale=0.3, bb=9 7 482 719]{Gd5Pt2In4_non-annealed_no_2021_03_powder_in_varnish_Hac=2_Oe_H_dc=0_Oe_M_vs_T_up_ZFC_20220305.eps}
		\includegraphics[scale=0.3, bb=9 7 482 719]{Tb5Pt2In4_non-annealed_no_2021_02_powder_in_varnish_Hac=2_Oe_H_dc=0_Oe_M_vs_T_up_ZFC_20220309.eps}
		\includegraphics[scale=0.3, bb=9 7 482 719]{Dy5Pt2In4_non-annealed_no_2021_06_powder_in_varnish_Hac=2_Oe_H_dc=0_Oe_M_vs_T_up_ZFC_20220316.eps}
		\includegraphics[scale=0.3, bb=9 7 482 719]{Ho5Pt2In4_non-annealed_no_2021_07_powder_in_varnish_Hac=2_Oe_H_dc=0_Oe_M_vs_T_up_ZFC_20220322.eps}  
		\includegraphics[scale=0.3, bb=9 7 482 719]{Er5Pt2In4_non-annealed_no_2021_13_powder_in_varnish_Hac=2_Oe_H_dc=0_Oe_M_vs_T_up_ZFC_20220326.eps}
		\includegraphics[scale=0.3, bb=9 7 482 719]{Tm5Pt2In4_non-annealed_no_2021_09_powder_in_varnish_Hac=2_Oe_H_dc=0_Oe_M_vs_T_up_ZFC_20220329.eps}
		\caption{\label{fig:AC_data}AC magnetic susceptibility of RE$_{5}$Pt$_2$In$_4$: (a) RE = Gd, (b) RE = Tb, (c) RE = Dy (d) RE = Ho, (e) RE = Er, and
		(f) RE = Tm taken at frequencies between 100~Hz and 5000~Hz. $\chi'$ and $\chi$ refer to the real and imaginary components, respectively.}
	\end{figure*}
	
	\subsection{Magnetic properties}
	\subsubsection{\label{DC_and_AC_magnetic_susceptibility_data}DC and AC magnetic susceptibility data}

The results of DC magnetic measurements of RE$_{5}$Pt$_2$In$_4$ (RE = Gd--Tm) are presented in Fig.~\ref{fig:Reciprocal_X_vs_T}
and summarized in Table~\ref{tbl:DC_AC_TC_TN_Tt_data}. At a low magnetic field of 50~Oe, all compounds show transitions from para- to magnetically ordered state
with decreasing temperature, namely, for RE = Gd, Tb, Ho and Er a rapid increase of magnetic susceptibility, characteristic of
para- to ferromagnetic transition, is visible, while for RE = Tm a maximum typical of para- to antiferromagnetic transition is
found (see the upper insets in Figs.~\ref{fig:Reciprocal_X_vs_T}a-f). The case of Dy$_{5}$Pt$_2$In$_4$ is more complicated as
the ferromagnetic transition at $T_C = 80$~K is preceded by an intermediate antiferromagnetic state manifesting itself by
a maximum at $T_N = 93$~K. Below the critical temperatures of magnetic ordering, a number of additional magnetic transitions
are detected from either additional maxima or inflection points. The magnetic transition temperatures are listed in
Table~\ref{tbl:DC_AC_TC_TN_Tt_data}. Large discrepancies between the ZFC and FC curves,
visible especially for RE = Gd--Ho below the corresponding Curie temperatures, are magnetic domains-related effects,
indicating presence of a ferromagnetic component of the magnetic order. The discrepancy found for RE = Tm,
both below and above the N\'eel temperature of 4.1~K, can be attributed to a very small amount of ferromagnetic
impurity phase, whose content is too small to be detectable by other experimental techniques like X-ray powder diffraction (XRD).

The maxima visible in the AC magnetic susceptibility vs. temperature curves (see Fig.~\ref{fig:AC_data}) coincide with the
magnetic transition temperatures derived from the DC magnetic data (see Table~\ref{tbl:DC_AC_TC_TN_Tt_data}). 
Some discrepancies between the DC and AC data can be attributed to different experimental conditions, namely,
the AC data have been collected under the oscillating field of 2~Oe and no applied DC field, while the DC data
have been recorded under a constant field of 50~Oe. Although relatively low, the latter field may influence
temperatures of magnetic transition sensitive to the applied magnetic field. It is worth noting that for
Dy$_{5}$Pt$_2$In$_4$, the N\'eel temperature derived from the DC data coincides with a distinct maximum observed
in $\chi'_{ac}$ at 93~K (there is no anomaly at this temperature in $\chi''_{ac}$), while the Curie point of 80~K is accompanied by maxima in
both $\chi'_{ac}$ and $\chi''_{ac}$.

	\begin{table*} 
		\begin{scriptsize}
			\begin{flushleft}
				\normalsize
				\caption{\label{tbl:DC_AC_TC_TN_Tt_data}Parameters characterizing magnetic order: $T_{C}$ (Curie temperature), $T_{N}$ (N\'eel temperature),
				$T_{t}$ (temperatures of additional anomalies), $\theta_{p}$ (paramagnetic Curie temperature), $\mu_{eff}$
				(effective magnetic moment), $\mu[\mu_B]$ (magnetic moment in the ordered state), $H_{cr}[kOe]$ (the critical
				field) and $H_{coer}[kOe]$ (the coercivity field) for RE$_{5}$Pt$_2$In$_4$ (RE 
				= Gd--Tm), as derived from DC and/or AC magnetometric measurements. The $i$ and $m$
				indices indicate whether the transition temperature corresponds to an inflection point or to a maximum in the
				$\chi(T)$ curve, respectively. The transition temperatures are calculated from the ZFC data, except few cases
				based on the FC data and marked explicitly by an additional index $f$.}
			\end{flushleft}
			\vspace{-0.2 cm}
			\footnotesize
			\tabcolsep=0cm
			\begin{tabular*}{0.99\textwidth}{@{\extracolsep{2pt}}lllllllllllllllll}
				\hline
				RE & \multicolumn{3}{c}{$T_{C}$[K]} &  \multicolumn{3}{c}{$T_{N}$[K]} & \multicolumn{3}{c}{$T_{t}$[K]} & $\theta_{p}[K]$ & \multicolumn{2}{c}{\textbf{$\mu_{eff}[\mu_B]$}} & \multicolumn{2}{c}{$\mu[\mu_B]$}  & $H_{cr}[kOe]$ & $H_{coer}[kOe]$ \\
				\cline{2-4} \cline{5-7} \cline{8-10} \cline{12-13} \cline{14-15}
				\multicolumn{1}{l}{} & \multicolumn{1}{l}{$\chi_{dc}$} & \multicolumn{1}{l}{$\chi'_{ac}$} & \multicolumn{1}{l}{$\chi''_{ac}$} & \multicolumn{1}{l}{$\chi_{dc}$} &
				\multicolumn{1}{l}{$\chi'_{ac}$} & \multicolumn{1}{l}{$\chi''_{ac}$} &
				\multicolumn{1}{l}{$\chi_{dc}$} & \multicolumn{1}{l}{$\chi'_{ac}$} & \multicolumn{1}{l}{$\chi''_{ac}$} & \multicolumn{1}{l}{} & \multicolumn{1}{l}{Exp.} & \multicolumn{1}{l}{Theor.} & \multicolumn{1}{l}{Exp.} & \multicolumn{1}{l}{Theor.} & \multicolumn{1}{l}{} & \multicolumn{1}{l}{}\\
				\hline
				Gd & 76$^{i}$ & 74$^{m}$ & 73$^{m}$ &  & &  &  22.5$^{i}$  &  &  & +72 & 7.87 & 7.94 & 3.86 & 7.00 & T = 1.9 K: 0.16 & 0.11\\
				&  &  &  &  &  &  &  &  &  &  &  &  &  &  &  T = 40.0 K: 0.05 & 0.02\\
				&  &  &  &  &  &  &  &  &  &  &  &  &  &  &  T = 68.0 K: 0.02 & \\
				Tb & 108$^{i}$ & 106$^{m}$ & 104$^{m}$ &  &  & & 83$^{i}$ & & & +48 & 9.55 & 9.72 & 4.34 & 9.00 & T = 1.9 K: 44 & 23.7\\
				&  &  &  &  &  &  &  &  &  &  &  &  &  &  &  T = 20.0 K: 1.7 & 1.9\\
				&  &  &  &  &  &  &  &  &  &  &  &  &  &  &  T = 100.0 K: 0.03 & \\
				Dy & 80$^{i}$ & 84$^{m}$ & 80$^{m}$  & 93$^{m}$ & 93$^{m}$ &  & 19.6$^{m,f}$, 71$^{i}$ & 14.1$^{m}$ &  & +24.6 & 10.61 & 10.65 & 5.74 & 10.00 & T = 1.9 K: 24.3, 45 & 22.1\\
				&  &  &  &  &  &  &  &  &  &  &  &  &  &  &  T = 12.0 K: 13.5 & 11.5\\
				&  &  &  &  &  &  &  &  &  &  &  &  &  &  &  T = 25.0 K: 2.2 & 2.4\\
				&  &  &  &  &  &  &  &  &  &  &  &  &  &  &  T = 75.0 K: 0.1 & 0.03\\
				&  &  &  &  &  &  &  &  &  &  &  & &  &  &  T = 87.0 K: 0.5 & \\
				Ho & 23.5$^{i}$ & 22.7$^{m}$ & 22.1$^{m}$ &  &  &  & 9.1$^{m,f}$, 13.5$^{i}$ &  8.1$^{m}$ &  & +10.1 & 10.62 & 10.61 & 6.98 & 10.00 &  T = 1.9 K: 2.9 & 1.4\\
				&  &  &  &  &  &  &  &  &  &  &  &  &  &  & T = 9.0 K: 0.22 & 0.23\\
				&  &  &  &  &  &  &  &  &  &  &  &  &  &  &  T = 15.0 K: 0.08 & 0.03\\
				&  &  &  &  &  &  &  &  &  &  &  &  &  &  & T = 20.0 K: 0.02 & \\
				Er & 12.6$^{i}$ & 12.1$^{m}$ & 11.9$^{m}$ &  &  &  & 6.9$^{i}$, 8.8$^{i}$ & & 7.1$^{m}$ & +11.1 & 9.44 & 9.59 & 3.24 & 9.00 & T = 1.9 K: 0.96 & 0.3\\
				&  &  &  &  &  &  &  &  &  &  &  &  &  &  & T = 7.0 K: 0.21 & 0.09\\
				&  &  &  &  &  &  &  &  &  &  &  &  &  &  & T = 9.5 K: 0.08 & 0.02\\
				Tm &   & & & 4.1$^{m}$ & 4.2$^{m}$ & 4.2$^{m}$ &  &  &  & +4.6 & 7.45 & 7.57 & 3.54 & 7.00 & T = 1.9 K: 9.2 & 0.03\\
				&  &  &  &  &  &  &  &  &  &  &  &  &  &  &  T = 3.0 K: 7.6 & 0.04\\
				&  &  &  &  &  &  &  &  &  &  &  &  &  &  &  T = 4.0 K: 0.07 & 0.02\\ \hline
			\end{tabular*}
			\\
			$i$ – inflection point; $m$ – maximum; $f$ – determined from the FC curve
		\end{scriptsize}
	\end{table*}

The reciprocal magnetic susceptibility curves, collected at H = 1~kOe, become linear at high-enough temperatures as predicted by the Curie-Weiss law~\cite{kittel2004introduction}:

	\setcounter{equation}{0}
	\renewcommand{\theequation}{\arabic{equation}}
	\begin{equation}
		\chi= \frac{C}{T-\theta_p} 
	\end{equation}

\noindent where $C$ is the Curie constant related to the effective magnetic moment ($\mu_{eff}$), while $\theta_p$ is a paramagnetic Curie temperature.
The values of $\mu_{eff}$ and $\theta_p$, as derived from fitting the linear dependence to $\chi_{dc}^{-1}(T)$ at high temperature regions, are
listed in Table~\ref{tbl:DC_AC_TC_TN_Tt_data}. The values of $\mu_{eff}$ are very close to those predicted for the free RE$^{3+}$ ions.
It is worth noting that for all investigated compounds, the determined paramagnetic Curie temperatures are positive, indicating the predominant role of the ferromagnetic interactions.

	\subsubsection{Magnetization}

The lower insets in Figs.~\ref{fig:Reciprocal_X_vs_T}a-f show magnetization curves for RE$_{5}$Pt$_2$In$_4$ (RE = Gd-Tm) taken
in applied magnetic fields up to 90~kOe at a number of selected temperatures. The shape of the curves, especially those collected
at lower temperatures, testifies to the coexistence of ferro- and antiferromagnetic components of the magnetic structures as both
metamagnetic transitions as well as magnetic hysteresis are observed. The critical fields, corresponding to the metamagnetic
transitions, have been determined from inflection points in the primary magnetization curves, i.e. the curves recorded directly after
the zero-field cooling (ZFC) procedure. The values of critical fields derived from experimental data are listed in Table~\ref{tbl:DC_AC_TC_TN_Tt_data}.
The appearance of metamagnetic transition confirms the existence of an antiferromagnetic component,
which is suppressed by application of a high-enough external magnetic field. The ferromagnetic component of the magnetic structure
manifests itself in non-zero coercivity field. The observed values of coercivity fields are also reported in Table~\ref{tbl:DC_AC_TC_TN_Tt_data}.

It is worth noting that both the Tb- and Dy-based compounds show very high values of critical and coercivity fields at 1.9~K,
testifying to stronger magnetic interactions than in the other investigated compounds. Such a result coincides with the highest
ordering temperatures observed for RE = Tb and Dy.

The case of Dy$_{5}$Pt$_2$In$_4$ is very special as the critical field initially decreases with increasing temperature, reaching
0.1~kOe at 75~K, and afterward increases again to 0.5~kOe at 87~K. Such a result is in agreement with the susceptibility data
(see subsection~\ref{DC_and_AC_magnetic_susceptibility_data}), which suggest suppression of the ferromagnetic component of
the Dy$_{5}$Pt$_2$In$_4$ magnetic structure at $T_C = 80$~K with further development of the antiferromagnetic one, which exists
up to $T_N = 93$~K.

The determined values of the magnetic moments in the ordered state, as derived from magnetization data taken at 1.9~K and 90~kOe,
are significantly lower than the values expected for free RE$^{3+}$ ions (for example, for Tb$_{5}$Pt$_2$In$_4$ it is 4.34~$\mu_B$,
which is only 48~\% of the theoretical value for Tb$^{3+}$ which equals 9.00~$\mu_B$ -- for other compounds, see Table~\ref{tbl:DC_AC_TC_TN_Tt_data}).
Nevertheless, one should take into account that the magnetization curves collected at 1.9~K are far away from saturation
even at a relatively high magnetic field of 90~kOe.

	\subsubsection{Magnetocaloric effect}

	\paragraph{Magnetic entropy change and temperature averaged magnetic entropy change (TEC)\\} 
	
	\begin{figure*}
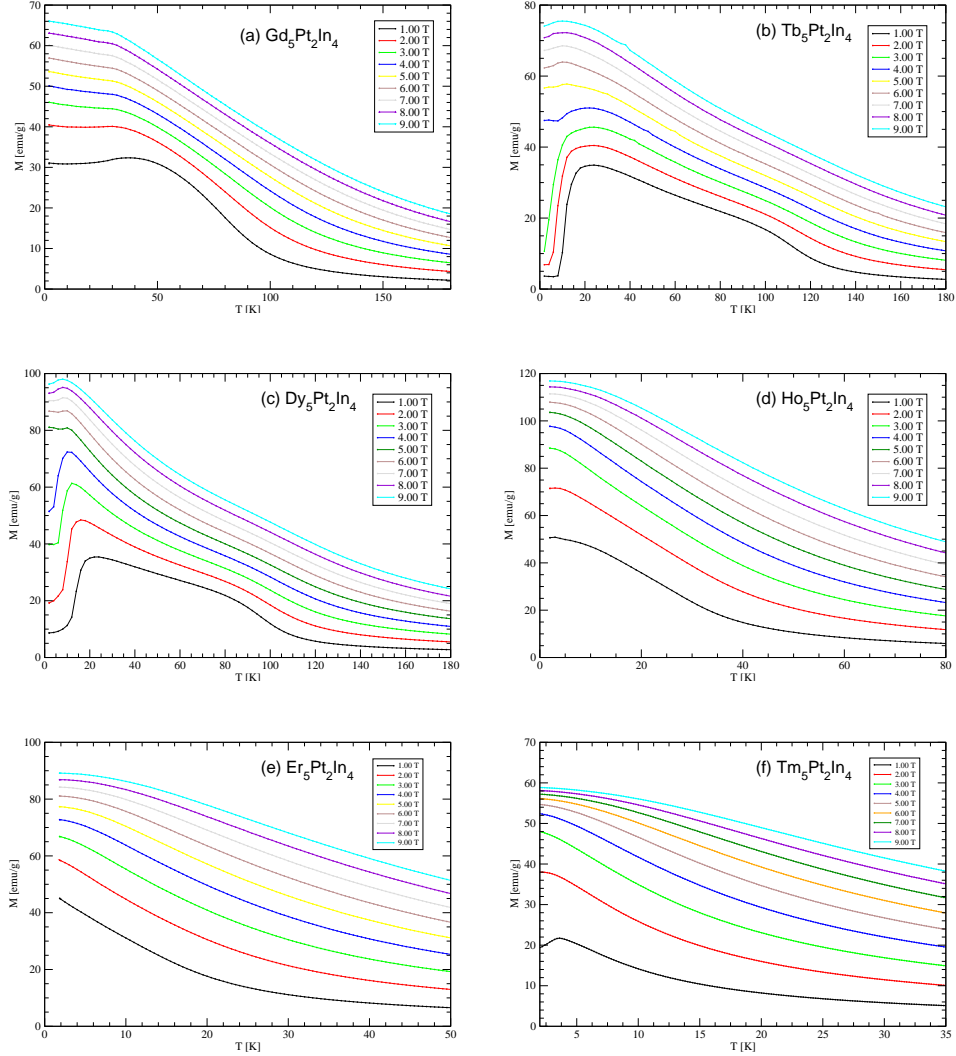

		\centering 
		\includegraphics[scale=0.25, bb=16 21 750 573]{Gd5Pt2In4_non-annealed_no_2021_03_powder_in_varnish_moment_vs_temp_up_ZFC.eps}
		\includegraphics[scale=0.25, bb=16 21 750 573]{Tb5Pt2In4_non-annealed_no_2021_02_powder_in_varnish_moment_vs_temp_up_ZFC.eps}
		\includegraphics[scale=0.25, bb=16 21 750 573]{Dy5Pt2In4_non-annealed_no_2021_06_powder_in_varnish_moment_vs_temp_up_ZFC.eps}
		\includegraphics[scale=0.25, bb=16 21 750 573]{Ho5Pt2In4_non-annealed_no_2021_07_powder_in_varnish_moment_vs_temp_up_ZFC.eps}
		\includegraphics[scale=0.25, bb=16 21 750 573]{Er5Pt2In4_non-annealed_no_2021_13_powder_in_varnish_moment_vs_temp_up_ZFC.eps}
		\includegraphics[scale=0.25, bb=16 21 750 573]{Tm5Pt2In4_non-annealed_no_2021_09_powder_in_varnish_moment_vs_temp_up_ZFC.eps}
		\caption{\label{fig:M_vs_T}Magnetization vs temperature curves under various magnetic flux density values up to 9~T for
		RE$_{5}$Pt$_2$In$_4$: (a) RE = Gd, (b) RE = Tb, (c) RE = Dy (d) RE = Ho, (e) RE = Er, and (f) RE = Tm.}
	\end{figure*}
	
	\begin{figure*}
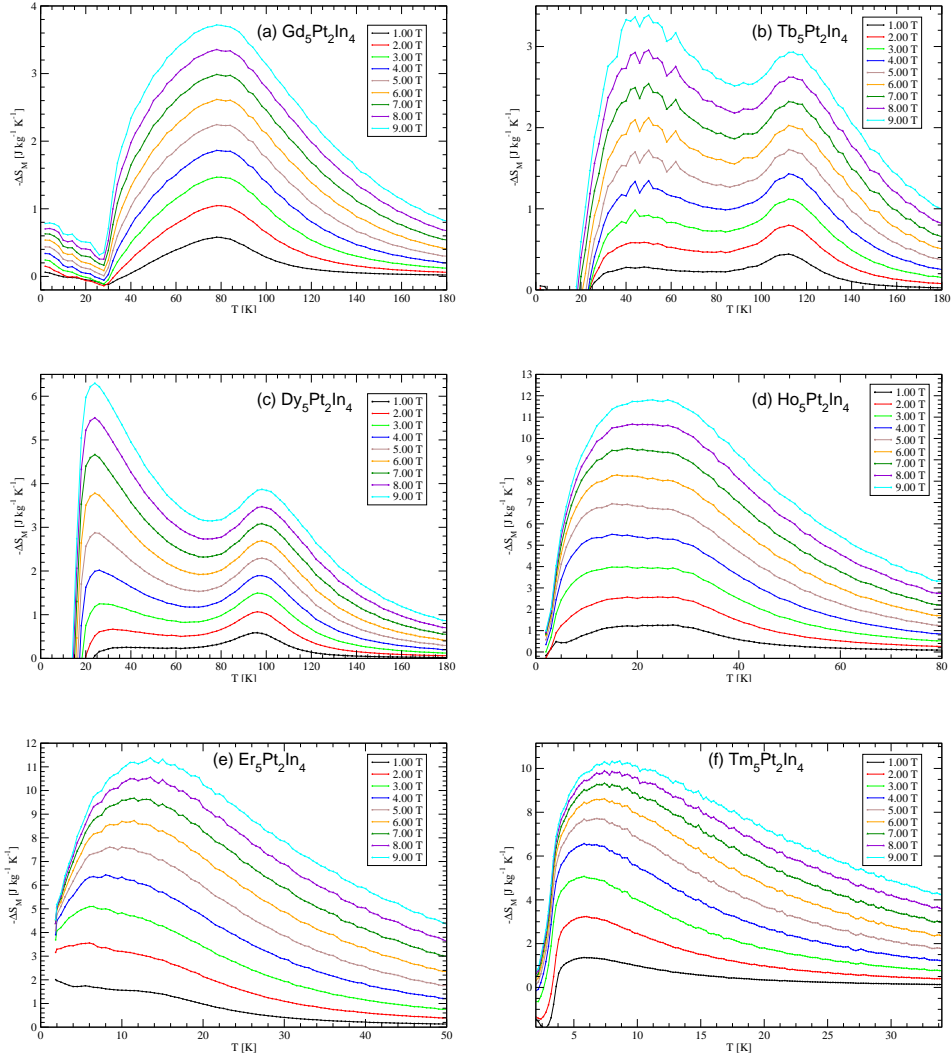

		\centering
		\includegraphics[scale=0.25, bb=16 21 750 573]{Gd5Pt2In4_non-annealed_no_2021_03_powder_in_varnish_entr_vs_temp_up_ZFC.eps}
		\includegraphics[scale=0.25, bb=16 21 750 573]{Tb5Pt2In4_non-annealed_no_2021_02_powder_in_varnish_entr_vs_temp_up_ZFC.eps}
		\includegraphics[scale=0.25, bb=16 21 750 573]{Dy5Pt2In4_non-annealed_no_2021_06_powder_in_varnish_entr_vs_temp_up_ZFC.eps}
		\includegraphics[scale=0.25, bb=16 21 750 573]{Ho5Pt2In4_non-annealed_no_2021_07_powder_in_varnish_entr_vs_temp_up_ZFC.eps}
		\includegraphics[scale=0.25, bb=16 21 750 573]{Er5Pt2In4_non-annealed_no_2021_13_powder_in_varnish_entr_vs_temp_up_ZFC.eps}
		\includegraphics[scale=0.25, bb=16 21 750 573]{Tm5Pt2In4_non-annealed_no_2021_09_powder_in_varnish_entr_vs_temp_up_ZFC.eps}
		\caption{\label{fig:magn_entr_vs_T}Temperature dependence of the magnetic entropy change $-\Delta S_M^{max}$, as
		derived from the $M(H,T)$ data, under various magnetic flux density changes $\Delta \mu_{0} H$ up to 0--9~T, for
		RE$_{5}$Pt$_2$In$_4$: (a) RE = Gd, (b) RE = Tb, (c) RE = Dy (d) RE = Ho, (e) RE = Er, and (f) RE = Tm.}
	\end{figure*}
	
	\begin{figure*}
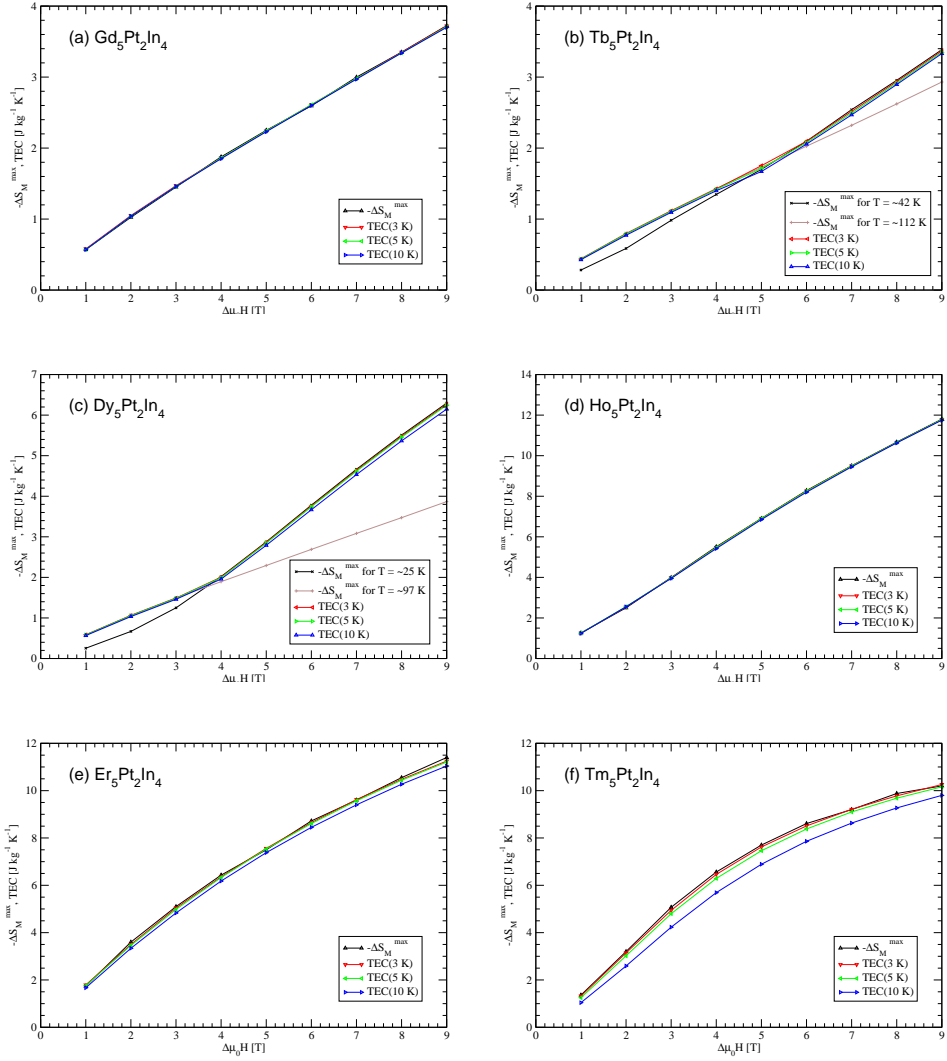

		\centering
		\includegraphics[scale=0.25, bb=16 21 750 573]{Magn_entropy-TECvsDnu0H-Gd5Pt2In4.eps}
		\includegraphics[scale=0.25, bb=16 21 750 573]{Magn_entropy-TECvsDnu0H-Tb5Pt2In4.eps}
		\includegraphics[scale=0.25, bb=16 21 750 573]{Magn_entropy-TECvsDnu0H-Dy5Pt2In4.eps}
		\includegraphics[scale=0.25, bb=16 21 750 573]{Magn_entropy-TECvsDnu0H-Ho5Pt2In4.eps}
		\includegraphics[scale=0.25, bb=16 21 750 573]{Magn_entropy-TECvsDnu0H-Er5Pt2In4.eps}
		\includegraphics[scale=0.25, bb=16 21 750 573]{Magn_entropy-TECvsDnu0H-Tm5Pt2In4.eps}
		\caption{\label{fig:tec}The values of $-\Delta S_M^{max}$, TEC(3 K), TEC(5 K), and TEC(10 K) under various magnetic
		flux density changes $\Delta \mu_{0}H$ up to 0--9 T for RE$_{5}$Pt$_2$In$_4$: (a) RE = Gd, (b) RE = Tb,
		(c) RE = Dy, (d) RE = Ho, (e) RE = Er, and (f) RE = Tm, respectively. For RE = Tb and Dy, the magnetic entropy changes
		corresponding to the low- and high-temperature maxima are indicated by black and brown lines, respectively.}
	\end{figure*}

Considering the magnetic entropy ($S_M$) as a function of magnetic flux density ($\mu_{0}H$) and temperature ($T$), its differential $\mathrm{d}S_M$ can be expressed as:

	\begin{equation}
	\mathrm{d}S_M = \left(\frac{\partial S_M}{\partial {(\mu_{0}H)}}\right)_{T} \mathrm{d}(\mu_{0}H) + \left(\frac{\partial S_M}{\partial  T}\right)_{(\mu_{0}H)} \mathrm{d}T
	\label{eqn:ds1}
	\end{equation}

Taking into account one of the Maxwell’s relations, namely that of
$\left(\frac{\partial S_M}{\partial {(\mu_{0}H})}\right)_{T} = \left(\frac{\partial M}{\partial T}\right)_{(\mu_{0}H)}$,
and inserting it to Eq.~\ref{eqn:ds1}, leads to:

	\begin{equation}
	\mathrm{d}S_M = \left(\frac{\partial M}{\partial T} \right)_{(\mu_{0}H)} \mathrm{d}(\mu_{0}H) + \left(\frac{\partial S_M}{\partial T} \right)_{(\mu_{0}H)} \mathrm{d}T
	\label{eqn:ds2}
	\end{equation} 

\noindent where $M$ denotes magnetization.

By applying the isothermal condition ($\mathrm{d}T = 0$) and integrating Eq.~\ref{eqn:ds2} over magnetic flux density, one gets:

\begin{equation}\Delta S_M(T,\Delta\mu_{0}H) = \int_{0}^{\mu_{0}H_{max}}  
               \left( \frac{\partial M (\mu_{0}H,T)}{\partial T} \right)_{(\mu_{0}H)}  \,\mathrm{d}\mu_{0}H
\label{eqn:dSm}
\end{equation}

\noindent where $\Delta\mu_{0}H$ is a change of the magnetic flux density (usually calculated with respect to the initial
flux equal to zero), while $\left( \frac{\partial M (\mu_{0}H,T)}{\partial T} \right)_{(\mu_{0}H)}$ is a derivative
of magnetization over temperature at fixed magnetic flux density of $\mu_{0}H$. Details of mathematical formalism
related to the magnetocaloric effect can be found for example in the book by Tishin and Spichkin~\cite{tishin2003magnetocaloric}.

Fig.~\ref{fig:M_vs_T} shows the magnetization vs. temperature $M(T)$ curves collected at a number of fixed magnetic flux density
values up to 9~T for the RE$_{5}$Pt$_2$In$_4$ (RE = Gd--Tm) compounds. Based on the above data, the corresponding magnetic entropy
changes have been calculated using Eq.~\ref{eqn:dSm}. The results are presented in Fig.~\ref{fig:magn_entr_vs_T}. Maximum entropy
changes around corresponding magnetic transition temperatures reach 3.7, 3.4, 6.3, 11.8, 11.4, and 10.2~J$\cdot$kg$^{-1}\cdot$K$^{-1}$
under magnetic flux density change of 0--9~T for RE = Gd, Tb, Dy, Ho, Er, and Tm, respectively. The values for other magnetic flux
density changes are listed in Table~\ref{tbl:Tc_DeltaSM_RCP}. The cases of Tb$_{5}$Pt$_2$In$_4$ and Dy$_{5}$Pt$_2$In$_4$ need an
extra comment, as two distinct maxima are found in the magnetic entropy change vs. temperature plots (see Figs.~\ref{fig:magn_entr_vs_T}b
and \ref{fig:magn_entr_vs_T}c). The high-temperature maximum ($\sim$110~K for RE = Tb and $\sim$95~K for RE = Dy), which dominates
for low magnetic flux density changes, corresponds to the transition from para- to ferromagnetic state. The low-temperature maximum
($\sim$45~K for RE = Tb and $\sim$25~K for RE = Dy), which dominates for high magnetic flux density changes, indicates an extra
transformation of the magnetic structure. The magnetic entropy changes corresponding to the low- and high-temperature maxima
are shown separately in Figs.~\ref{fig:magn_entr_vs_T}b and \ref{fig:magn_entr_vs_T}c.

The temperature averaged magnetic entropy change (TEC) is another parameter used for evaluating MCE. TEC is defined by
the following formula ~\cite{griffith2018material-based}:

\begin{align} 
\MoveEqLeft TEC(\Delta T_\mathrm{lift},\Delta\mu_{0}H) =
\nonumber \\  &\frac{1}{\Delta T_\mathrm{lift}} \max_{T_\mathrm{mid}}
  \Bigg\{\int_{T_\mathrm{mid}-\frac{\Delta T_\mathrm{lift}}{2}}^{T_\mathrm{mid}+\frac{\Delta T_\mathrm{lift}}{2}} \Delta S_{\mathrm{M}}(T,\Delta\mu_{0}H)\,\mathrm{d}T\Bigg\}
\label{eqn:tec}
\end{align}

\noindent where $T_{mid}$ is the center temperature of the temperature span ${\Delta T_{lift}}$. The value of $T_{mid}$ is determined
by finding the one that maximizes the integral appearing in Eq.~\ref{eqn:tec}. The TEC values of RE$_{5}$Pt$_2$In$_4$ (RE = Gd--Tm),
calculated for the temperature spans of 3, 5 and 10~K, are presented in Fig.~\ref{fig:tec}. It is worth noting that the TEC values
for RE = Tb and Dy are always related to the dominating maximum, i.e. to the high-temperature maximum for the low magnetic flux density
changes and the low-temperature one for the high magnetic flux density changes (see Figs.~\ref{fig:tec}b and \ref{fig:tec}c).

\bigskip	

\paragraph{Relative Cooling Power (RCP) and Refrigerant Capacity (RC)\\}

Besides $\Delta S_M^{max}$ and TEC, the relative cooling power (RCP)~\cite{gschneidner_pecharsky2000magnetocaloric_materials} and refrigerant
capacity (RC)~\cite{wood_potter1985general_analysis} are another important parameters that can be used to assess the MCE performance.
RCP is defined by the equation:

\begin{equation}
 RCP = -\Delta S_M^{max} \times \delta T_{FWHM}
\label{eqn:rcp}
\end{equation}

\noindent where $\delta T_{FWHM}$ is the full width at half maximum of the entropy change vs. temperature curve. RC is calculated from the
following integral:

\begin{equation}
 RC = \int_{T_{1}}^{T_{2}} |\Delta S_M| \,\mathrm{d}T
\label{eqn:rc}
\end{equation} 

\noindent where $T_1$ and $T_2$ denote the lower and upper limits of the FWHM temperature range, respectively.

\begin{table*}

\begin{footnotesize}
  \begin{flushleft}
  \normalsize
\caption{\label{tbl:Tc_DeltaSM_RCP}\ The temperatures of maximum entropy change, maximum entropy changes $-\Delta S_M^{max}$
together with RCP and RC values under magnetic flux density changes $\Delta \mu_0 H$ = 0--2 T, 0--5 T, 0--7 T and 0--9 T for the RE$_{5}$Pt$_2$In$_4$ (RE 
= Gd--Tm) and isostructural compounds.\\}
  \end{flushleft}
  \vspace{-0.2 cm}
  \footnotesize
  \begin{tabular*}{0.99\textwidth}{@{\extracolsep{\fill}}lllllllllllllll}
    \hline
    Materials & Temp. for $-\Delta S_{\mathrm{M}}^{\mathrm{max}}$ [K] & \multicolumn{4}{c}{$-\Delta S_{\mathrm{M}}^{\mathrm{max}}$[J$\cdot$kg$^{-1}\cdot$K$^{-1}$]} & \multicolumn{4}{c}{RCP [J$\cdot$kg$^{-1}$]} & \multicolumn{4}{c}{RC [J$\cdot$kg$^{-1}$]} & Ref.\\
\cline{3-6} \cline{7-10} \cline{11-14}  
\multicolumn{1}{l}{} & \multicolumn{1}{l}{} & \multicolumn{1}{l}{0--2 T} & \multicolumn{1}{l}{0--5 T} & \multicolumn{1}{l}{0--7 T} & \multicolumn{1}{l}{0--9 T} &
	\multicolumn{1}{l}{0--2 T} & \multicolumn{1}{l}{0--5 T} & \multicolumn{1}{l}{0--7 T} & \multicolumn{1}{l}{0--9 T} & \multicolumn{1}{l}{0--2 T} & \multicolumn{1}{l}{0--5 T} & \multicolumn{1}{l}{0--7 T} & \multicolumn{1}{l}{0--9 T} & \\
    \hline
 Dy$_{5}$Ni$_2$In$_4$ & 19 & 1.8 & 3.6 & 4.7 & - & 49 & 178 & 286 & - & 37 & 130 & 209 & - & \citep{zhang2018investigation}\\
 Ho$_{5}$Ni$_2$In$_4$ & 103 & 2.6 & 7.1 & 10.1 & - & 84 & 298 & 458 & - & 66 & 234 & 352 & - & \citep{zhang2018investigation}\\
 Er$_{5}$Ni$_2$In$_4$ & 20 & 3.3 & 7.7 & 10.2 & - & 71 & 248 & 377 & - & 52 & 180 & 273 & - & \citep{zhang2018investigation}\\
 Gd$_{5}$Pt$_2$In$_4$ & 78 & 1.0 & 2.2 & 3.0 & 3.7 & 58 & 172 & 261 & 359 & 48 & 139 & 209 & 290 & this work\\
 Tb$_{5}$Pt$_2$In$_4$ & 45, 110 & 0.8 & 1.7 & 2.5 & 3.4 & 82 & 198 & 303 & 428 & 57 & 165 & 248 & 340 & this work\\
 Dy$_{5}$Pt$_2$In$_4$ & 25, 95 & 1.1 & 2.9 & 4.7 & 6.3 & 45 & 290 & 250 & 363 & 33 & 201 & 180 & 263 & this work\\
 Ho$_{5}$Pt$_2$In$_4$ & 23 & 2.5 & 6.9 & 9.5 & 11.8 & 94 & 302 & 451 & 607 & 82 & 254 & 373 & 495 & this work\\
 Er$_{5}$Pt$_2$In$_4$ & 14 & 3.6 & 7.5 & 9.6 & 11.4 & 79 & 218 & 328 & 434 & 63 & 175 & 256 & 341 & this work\\
 Tm$_{5}$Pt$_2$In$_4$ & 8 & 3.2 & 7.7 & 9.2 & 10.2 & 34 & 125 & 189 & 260 & 27 & 98 & 150 & 205 & this work\\
    \hline
  \end{tabular*}
\end{footnotesize}
\end{table*}

Table~\ref{tbl:Tc_DeltaSM_RCP} lists the determined values of RCP and RC for RE$_{5}$Pt$_2$In$_4$ (RE = Gd--Tm)
under selected changes of magnetic flux densities. It is worth noting that for Dy$_{5}$Pt$_2$In$_4$,
the RCP and RC values under magnetic flux density changes of 0--5~T are higher than those under
0--7~T. This unusual behavior is due to the fact that under $\Delta \mu_0 H$ = 0--5~T, the low- and high-temperature maxima
in $-\Delta S_M(T)$ are of similar heights and they overlap, leading to FWHM temperature interval being
significantly wider than the one at higher magnetic flux density changes where low-temperature maximum
dominates.

\section{Discussion}	
This work reports the results of X-ray, DC, and AC magnetic measurements for the RE$_{5}$Pt$_2$In$_4$ (RE = Gd-Tm) compounds. The X-ray
diffraction data confirm that RE$_{5}$Pt$_2$In$_4$ (RE = Gd-Tm) have an orthorhombic crystal structure of the Lu$_{5}$Ni$_2$In$_4$-type.
The crystal structure is a typical two-layered one with layers formed by the rare earth atoms $(z = 0)$ separated by layers containing
the remaining Pd and In atoms $(z = \frac{1}{2})$. The refined lattice parameters and atomic coordinates, presented in Table~\ref{tbl:crystallographic_data},
are in good agreement with the previously reported ones~\cite{zaremba2007}.

The results of magnetometric measurements reveal that all investigated compounds order magnetically with decreasing temperature. The shapes of the magnetic susceptibility vs. temperature curves (see Figs.~\ref{fig:Reciprocal_X_vs_T}, \ref{fig:AC_data}) suggest para- to ferromagnetic transition for RE = Gd, Tb, Ho and Er, while for RE = Tm a para- to antiferromagnetic transition is found.
The case of RE = Dy is more complicated as the ferromagnetic state is reached through an intermediate antiferromagnetic state
occuring in realtively narrow temperature range. Anomalies in the $\chi(T)$ curves, observed below the respective critical
temperatures of magnetic ordering for all compounds except RE = Tm, suggest presence of additional temperature-induced magnetic
transitions. Such a behavior has been observed in the isostructural
RE$_5$Ni$_2$In$_4$~\cite{tyvanchuk2010magnetic,provino2012crystal,GONDEK201210,szytula2013magnetic,SZYTULA2014149,Ritter_2015,zhang2018investigation}.
It can be attributed to complexity of the crystal structure as the rare earth atoms occupy three nonequivalent Wyckoff sites,
which leads to competition of different interactions. Complex magnetic properties manifest themselves also in coexistance
of the ferro- and antiferromagnetic contributions to the magnetic structure, as confirmed by the shapes of the magnetization
vs. applied magnetic field curves, which show both presence of coercivity fields characteristic of ferromagnetic order as well as
metamagnetic transitions characteristic of antiferromagnetic ordering (see the lower insets in Figs.~\ref{fig:Reciprocal_X_vs_T}a-f).
It has to be mentioned that the exact thermal evolution of the magnetic structures cannot be determined solely from the magnetic data, but further neutron diffraction studies are required.

\begin{figure}[ht!]
	\centering
	\includegraphics[width=0.5\textwidth]{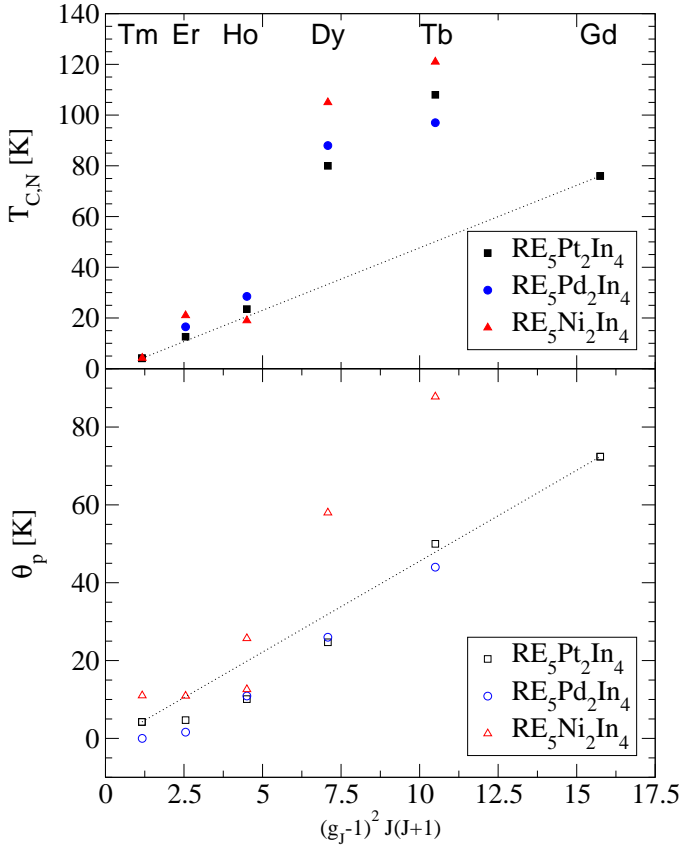}
	\caption{\label{fig:lande_factor_vs_Theta_P}Critical temperatures of magnetic ordering ($T_{C,N}$) along with paramagnetic Curie temperatures $\theta_p$ vs. de Gennes factor for RE$_{5}$T$_2$In$_4$ (RE = rare earth element, T = Ni, Pd, Pt). The data for T = Ni have been taken after~\cite{tyvanchuk2010magnetic,provino2012crystal,GONDEK201210,SZYTULA2014149,Ritter_2015}, while those for T = Pd after \cite{Baran2021}. The dotted lines indicate theoretical de Gennes scaling for RE$_{5}$Pt$_2$In$_4$ while taking the characteristic temperatures of Gd$_{5}$Pt$_2$In$_4$ as reference.}
\end{figure}

Reciprocal magnetic susceptibilities of RE$_{5}$Pt$_2$In$_4$ (RE = Tb-Tm) follow the Curie-Weiss law at high temperatures
(see Fig.~\ref{fig:Reciprocal_X_vs_T}).
The deviations from linearity, as observed at lower temperatures, indicate influence of the crystalline electric field (CEF).
The determined Curie temperatures are positive revealing that the ferromagnetic interactions are dominant.
The obtained values of the effective magnetic moments are close to those redicted for the free RE$^{3+}$ ions.
Small discrepancies between the experimental and theoretical values do not exceed systematic errors of the experiment.
Therefore the magnetism of RE$_{5}$Pt$_2$In$_4$ (RE = Tb-Tm) is strictly related to the rare earth magnetic moments,
while the magnetic moments of the remaining Pt and In elements are either zero or they are too small to be detectable
when accompanied by high rare earth moments.

According to the crystallographic data reported in Tables 1 and 4 in \cite{zaremba2007}, the rare earth interatomic distances
in RE$_{5}$Pt$_2$In$_4$ (RE = Tb-Tm) exceed 3.3~\AA. Therefore, they are high enough to exclude any direct interactions
and the indirect interactions of the RKKY-type are expected in this family of compounds. One of the predictions of the
RKKY theory is so called de Gennes scaling, which assumes direct proportionality between the critical temperature of
magnetic ordering and the de Gennes factor defined as $(g_J-1)^2J(J+1)$, where $g$ is a Land\'e factor, while 
$J$ is a total angular momentum of the RE$^{3+}$ rare earth ion. Figure~\ref{fig:lande_factor_vs_Theta_P}
presents the critical temperatures of magnetic ordering for RE$_{5}$T$_2$In$_4$ (T = Ni, Pd, Pt) plotted against
de Gennes factor. In addition, the paramagnetic Curie temperatures, which are also a measure of the strength
of magnetic interactions, are shown. The lack of theoretically predicted proportionality is yet another evidence
of significant influence of CEF on magnetic properties of RE$_{5}$T$_2$In$_4$ (RE = rare earth element, T = Ni, Pd, Pt).
It is worth noting that the critical temperature of magnetic ordering of a particular Ni-based compound is in the most cases
higher than those of its Pd- and Pt-based analogues. This finding coincides with increase of the interatomic distances due to
increasing number of the atomic number of the d-electron element (compare the lattice parameters for
T~=~Ni~\cite{zaremba1991crystal,TYVANCHUK2008878,provino2012crystal}, Pd~\cite{Sojka200890} and Pt~\cite{zaremba2007}).
Therefore, the increase of the interatomic distances leads to weaken the magnetic interactions.

\begin{figure*}
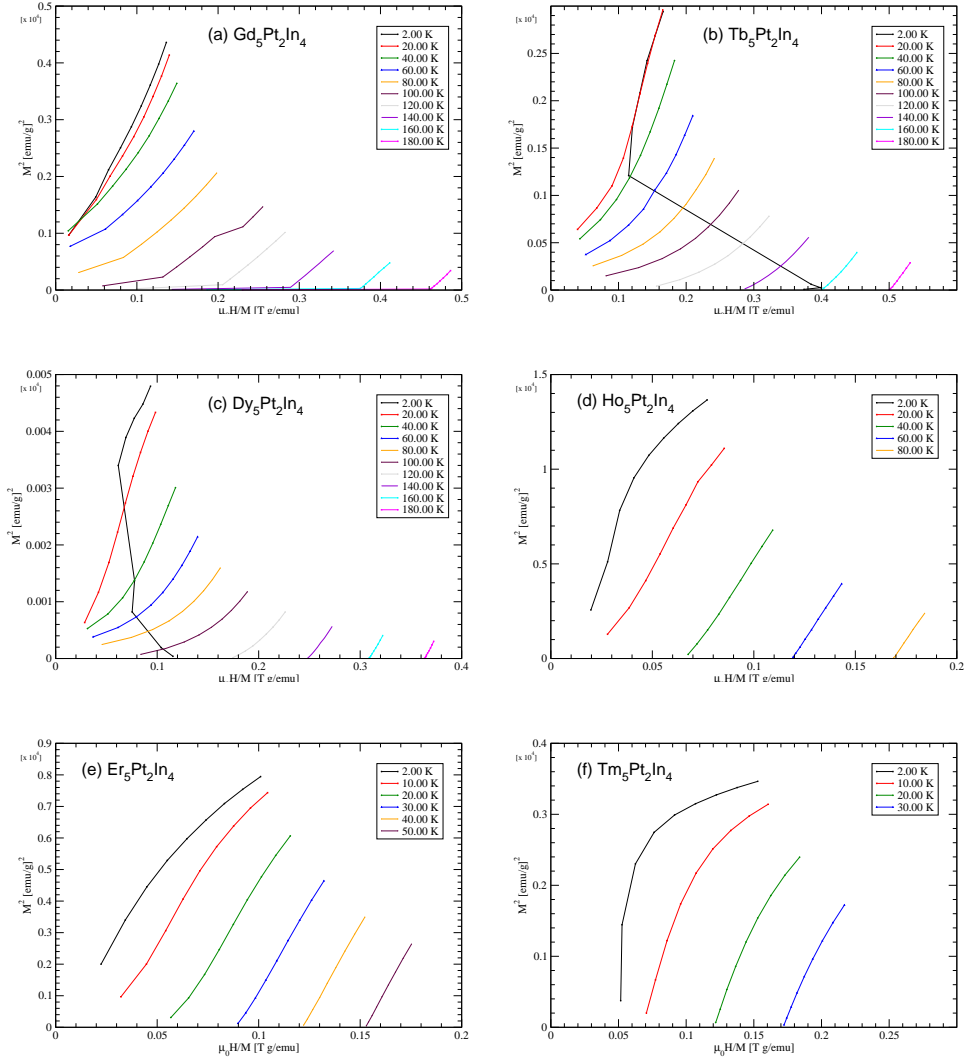

	\centering
	\includegraphics[scale=0.25, bb=16 21 750 573]{Gd5Pt2In4_non-annealed_no_2021_03_powder_in_varnish_M2-HperM_up_step_1T_202103_T.eps}
	\includegraphics[scale=0.25, bb=16 21 750 573]{Tb5Pt2In4_non-annealed_no_2021_02_powder_in_varnish_M2-HperM_up_step_1T_202104_T.eps}
	\includegraphics[scale=0.25, bb=16 21 750 573]{Dy5Pt2In4_non-annealed_no_2021_06_powder_in_varnish_M2-HperM_up_step_1T_202104_T.eps}
	\includegraphics[scale=0.25, bb=16 21 750 573]{Ho5Pt2In4_non-annealed_no_2021_07_powder_in_varnish_M2-HperM_up_step_1T_202105_T.eps}  
	\includegraphics[scale=0.25, bb=16 21 750 573]{Er5Pt2In4_non-annealed_no_2021_13_powder_in_varnish_M2-HperM_up_step_1T_202108_T.eps}
	\includegraphics[scale=0.25, bb=16 21 750 573]{Tm5Pt2In4_non-annealed_no_2021_09_powder_in_varnish_M2-HperM_up_step_1T_202111_T.eps}
	\caption{\label{fig:M2_vs_H_over_M}The plots of $M^{2}$ versus $\mu_{0} H/M$ at selected temperatures for RE$_{5}$Pt$_2$In$_4$:
	(a) RE = Gd, (b) RE = Tb, (c) RE = Dy (d) RE = Ho, (e) RE = Er and (f) RE = Tm.\\
	}
\end{figure*}

The type of magnetic transition (the first or second order) can be derived from shapes of the Arrott plots ($M^{2}$ vs. $\mu_{0} H/M$)
collected at selected temperatures~\cite{banerjee1964generalised_approach}. Positive slope of the Arrott curve corresponds to the
second order phase transitions (SOPT), while the negative slope to the first order phase transitions (FOPT).
Fig~\ref{fig:M2_vs_H_over_M} shows the Arrott plots for RE$_{5}$Pt$_2$In$_4$ (RE = Gd-Tm). For RE~=~Gd, Ho, Er and Tm only positive
slopes are found, indicating magnetic transition of the second order type. The situation is more complicated for RE~=~Tb and Dy, as
positive slopes characteristic of SOPT are found below the critical temperature of magnetic ordering, except the lowest temperature
of 2~K, where positive slope characteristic of FOPT is observed within limited range. This behavior is consistent with the shapes of
the magnetization vs. temperature curves (see Fig.~\ref{fig:M_vs_T}). For the low values of the applied magnetic field, magnetization
initially increases with decreasing temperature, but finally undergoes a sudden drop below 20~K. Such a behavior indicates that
the high-temperature ferro-/ferrimagnetic structure transforms into an antiferromagnetic one with decreasing temperature. According to
the shape of Arrott plots this phase transition is of the first order type (FOPT).

{
\renewcommand{\arraystretch}{1.0}
\begin{table*}
\caption{\label{tbl:magncal_perf}Comparison of the magnetocaloric performance under magnetic flux density change of 0-7~T
for RE$_{5}$Pt$_2$In$_4$ (RE = Gd-Tm), isostructural RE$_{5}$Ni$_2$In$_4$ (RE = Dy, Ho, Er) and other selected
ternary rare earth-based indides. $T_{cr}$ denotes critical temperature of the magnetic ordering (Curie or N\'eel temperature).\\
}
{\begin{tabular*}{\textwidth}{@{\extracolsep{\fill}}llllll}
    \hline 
Material & $T_{cr}$ [K] & $-\Delta S_{\mathrm{M}}^{\mathrm{max}}$ [J$\cdot$kg$^{-1}\cdot$K$^{-1}$] &
        RCP $[\mathrm{J}$$\cdot$$\mathrm{kg}^{-1}$] & RC $[\mathrm{J}$$\cdot$$\mathrm{kg}^{-1}$] & Ref.\\

\hline
Gd$_{5}$Pt$_2$In$_4$ & 76 & 3.0 & 261 & 209 & this work\\
Tb$_{5}$Pt$_2$In$_4$ & 108 & 2.5 & 303 & 248 & this work\\
Dy$_{5}$Pt$_2$In$_4$ & 93 & 4.7 & 250 & 180 & this work\\
Ho$_{5}$Pt$_2$In$_4$ & 23.5 & 9.5 & 451 & 373 & this work\\
Er$_{5}$Pt$_2$In$_4$ & 12.6 & 9.6 & 328 & 256 & this work\\
Tm$_{5}$Pt$_2$In$_4$ & 4.1 & 9.2 & 189 & 150 & this work\\
Dy$_{5}$Ni$_2$In$_4$ & 105 & 4.7 & 286 & 209 & \cite{zhang2018investigation}\\
Ho$_{5}$Ni$_2$In$_4$ & 31 & 10.1 & 458 & 352 & \cite{zhang2018investigation}\\
Er$_{5}$Ni$_2$In$_4$ & 21 & 10.2 & 377 & 273 & \cite{zhang2018investigation}\\
Gd$_{11}$Co$_4$In$_9$ & 86 & 10.95 & 538.1 & 405.9 & \cite{zhang2020structural}\\
Tb$_{11}$Co$_4$In$_9$ & 95$^{*}$ & 4.43 & 376.8 & 274.4 & \cite{Baran_R11Co4In9_Phase_Trans}\\
Dy$_{11}$Co$_4$In$_9$ & 37 & 4.66 & 213.9 & 165.9 & \cite{zhang2020structural}\\
Ho$_{11}$Co$_4$In$_9$ & 20 & 12.29 & 475.2 & 357.4 & \cite{zhang2020structural}\\
Er$_{11}$Co$_4$In$_9$ & 5.4$^{*}$ & 12.80 & 416.1 & 320.0 & \cite{Baran_R11Co4In9_Phase_Trans}\\
Gd$_{11}$Ni$_4$In$_9$ & 91 & 3.58 & 269.0 & 206.4 & \cite{ZHANG2021155863}\\
Dy$_{11}$Ni$_4$In$_9$ & 18 & 6.02 & 194.9 & 144.7 & \cite{ZHANG2021155863}\\
Ho$_{11}$Ni$_4$In$_9$ & 13.5 & 12.44 & 353.0 & 269.2 & \cite{ZHANG2021155863}\\
Gd$_{6}$Co$_{2.2}$In$_{0.8}$ & 76 & 11.84 & 814.23 & 633.55 & \cite{ZHANG2021107254}\\
Tb$_{6}$Co$_{2.2}$In$_{0.8}$ & 32 & 8.96 & 394.15 & 284.32 & \cite{ZHANG2021107254}\\
Dy$_{6}$Co$_{2.2}$In$_{0.8}$ & 50 & 9.59 & 517.01 & 389.77 & \cite{ZHANG2021107254}\\
Ho$_{6}$Co$_{2.2}$In$_{0.8}$ & 18 & 20.83 & 626.36 & 466.07 & \cite{ZHANG2021107254}\\
\hline
\end{tabular*}}
\vspace{-0.4cm} \flushleft {\noindent\footnotesize $^{*}$ after~\cite{baran2021crystal}}
\end{table*}
} 

Table~\ref{tbl:magncal_perf} contains comparison of the MCE performance under magnetic flux density change of 0-7~T
for RE$_{5}$Pt$_2$In$_4$ (RE = Gd-Tm), isostructural RE$_{5}$Ni$_2$In$_4$ (RE = Dy, Ho, Er) and other selected
ternary rare earth-based indides. For a selected rare earth element, the MCE performance of a member of the
RE$_{5}$Pt$_2$In$_4$ (RE = Gd-Tm) family of compounds shows similar performance to that of its RE$_{5}$Ni$_2$In$_4$
and RE$_{11}$T$_4$In$_9$ (T = Co, Ni) analogues. It is worth mentioning that the MCE performance of RE$_{5}$Pt$_2$In$_4$
(RE = Ho and Er), which is reported in this work, is quite high and comparable to that of the best known low-temperature magnetocaloric materials~\cite{
LI20231,Guo2022,XU2022118114,ZHANG2022100786,MA2021138,ZHANG20191173,LI2020153810,ZHANG2022117669,XU2021100470,LI2020354,ZHANG202266,
zhang2020structural,Gschneidner_et_al_2005_Recent_developments,lyubina2017magnetocaloric,ZHANG2021155863,zhang2018investigation,ZHANG2021107254}.
	
\section{Conclusions}
The RE$_{5}$Pt$_2$In$_4$ (RE = Gd-Tm) compounds crystallize in an orthorhombic crystal structure of the
the Lu$_{5}$Ni$_2$In$_4$-type (\textit{Pbam} space group) in which the rare earth atoms occupy three nonequivalent sites.
The compounds show complex magnetic properties and quite good MCE performance at low temperatures.
With decreasing temperature all investigated compounds undergo a transition to magnetically ordered
state, in the most cases followed by a cascade of extra magnetic transitions appearing below the respective critical
temperature of magnetic ordering. The rare earth atoms are found to possess magnetic moments, while the moments of the
remaining Pt and In atoms are absent or are too small to be detected while accompanied by the strong rare earth's moments.
The MCE performance of RE$_{5}$Pt$_2$In$_4$ (RE = Gd-Tm) is comparable to that of the best known low-temperature
magnetocaloric materials, especially while taking into account the compounds with RE = Ho and Er.

\section*{Declaration of Competing Interest}
The authors declare no conflict of interest.

\section*{Acknowledgements}
This research was supported in part by the Excellence Initiative --
Research University Program at the Jagiellonian University in Krak\'ow. The research was partially carried out with the equipment purchased thanks to the financial support of the European Regional Development Fund in the framework of the Polish Innovation Economy Operational Program (contract no.POIG.02.01.00-12-023/08).

	\bibliographystyle{elsarticle-num-names} 
	\bibliography{RE5Pt2In4_magn_effect_arXiv_20230425_AH.bib}
\end{document}